\documentclass[]{aastex631}
\pdfoutput=1
\usepackage{bm}
\usepackage{graphicx}
\usepackage{footnote}
\usepackage{color}
\usepackage{amsmath}
\usepackage{amsfonts}
\usepackage{mathtools}
\usepackage{multirow}
\newcommand{\mubold}{\bm{\mu}}
\newcommand\ltsima{$\; \buildrel <\over\sim \;$}
\newcommand\simlt{\lower.5ex\hbox{\ltsima}}
\newcommand\gtsima{$\; \buildrel >\over\sim \;$}
\newcommand\simgt{\lower.5ex\hbox{\gtsima}}

\newcommand\msun {M_\odot}

\shorttitle{}
\shortauthors{Bhattacharya et al}

\begin{document}

\title{OGLE-2012-BLG-0563: Relative Proper Motion Constraint from Keck Follow-up finds the correct Lens }


\author{Aparna Bhattacharya}
\affiliation{Department of Astronomy,
    University of Maryland, College Park, MD 20742, USA}
\affiliation{Code 667, NASA Goddard Space Flight Center, Greenbelt, MD 20771, USA}
\email{abhatta5@umd.edu}

\author{David~P. Bennett}
\affiliation{Department of Astronomy,
    University of Maryland, College Park, MD 20742, USA}
\affiliation{Code 667, NASA Goddard Space Flight Center, Greenbelt, MD 20771, USA}

\author{Jean~Philippe~Beaulieu}
\affiliation{UPMC-CNRS, UMR 7095, Institut d’Astrophysique de Paris, 98Bis Boulevard Arago, F-75014 Paris}
\affiliation{School of Physical Sciences, University of Tasmania, Private Bag 37 Hobart, Tasmania 7001 Australia} 

\author{Ian~A.~Bond}
\affiliation{Institute of Natural and Mathematical Sciences, Massey University, Auckland 0745, New Zealand}

\author{Kansuke~Nunota}
\affiliation{Department of Earth and Space Science,  Graduate School of Science, University of Osaka, Toyonaka, Osaka 560-0043, Japan}

\author{Jessica~R.~Lu}
\affiliation{University of California Berkeley, Berkeley, CA}

\author{Joshua~W.~Blackman}
\affiliation{School of Physical Sciences, University of Tasmania, Private Bag 37 Hobart, Tasmania 7001 Australia} 

\author{Andrew~A.~Cole}
\affiliation{School of Physical Sciences, University of Tasmania, Private Bag 37 Hobart, Tasmania 7001 Australia}

\author{Akihiko~Fukui}
\affiliation{Komaba Institute for Science, The University of Tokyo, 3-8-1 Komaba, Meguro, Tokyo 153-8902, Japan}
\affiliation{Instituto de Astrof\'{i}sica de Canarias (IAC), 38205 La Laguna, Tenerife, Spain}

\author{Jean~Baptiste~Marquette}
\affiliation{UPMC-CNRS, UMR 7095, Institut d’Astrophysique de Paris, 98Bis Boulevard Arago, F-75014 Paris}

\author{Cl{\'e}ment~Ranc} 
\affiliation{UPMC-CNRS, UMR 7095, Institut d’Astrophysique de Paris, 98Bis Boulevard Arago, F-75014 Paris}

\author{Natalia Rektsini}
\affiliation{UPMC-CNRS, UMR 7095, Institut d’Astrophysique de Paris, 98Bis Boulevard Arago, \affiliation{School of Physical Sciences, University of Tasmania, Private Bag 37 Hobart, Tasmania 7001 Australia}F-75014 Paris}

\author{Sean K. Terry} 
\affiliation{Code 667, NASA Goddard Space Flight Center, Greenbelt, MD 20771, USA}
\affiliation{Department of Astronomy,
    University of Maryland, College Park, MD 20742, USA} 
 
\author{Aikaterini Vandorou}
\affiliation{Code 667, NASA Goddard Space Flight Center, Greenbelt, MD 20771, USA}
\affiliation{Department of Astronomy,
    University of Maryland, College Park, MD 20742, USA}

\begin{abstract}
We present Keck/NIRC2 adaptive optics imaging results of planetary microlensing event OGLE-2012-BLG-0563. We find that the OGLE-2012-BLG-0563L planetary system consists
of a $1.12\pm 0.09 M_{\rm Jup}$ planet orbiting a $0.81\pm 0.04\msun$  host star at a distance 
of $5.49 \pm 0.56$ kpc from the Sun.  This event  has a lens-source separation of $23.28 \pm 0.48$ mas which is the smallest lens-source separation measured to date with Keck AO data. We find that the incorrect solution—which identifies the Northeast (NE) star as the source star—is favored if one only considers the discovery paper,  Fukui et al. (2015) solution or an unconstrained model.  However, by performing an image-constrained light curve model fit using the relative proper motion direction constraint from the Keck follow-up data, we demonstrate that the correct solution identifies the NE star as the lens star. The discrepancy in the unconstrained models is driven by systematics in the FTS and MOA B $\&$ C data,  alongside the impact of the relative proper motion direction constraints. This outcome offers two important lessons for future {\it Roman} data analysis: 1) thoroughly verify systematics across all passbands and identify the precise origin of $\chi^2$ differences, and 2) utilize image-constrained light curve modeling incorporating relative proper motion direction constraints to identify the correct lens. 

\end{abstract}

\section{Introduction}
\label{sec-intro}
Gravitational microlensing is unique in its ability to detect low mass exoplanets \citep{bennett96} beyond the 
snow line \citep{gouldloeb1992}, where the formation of giant planets is thought to be most efficient
\citep{pollack96,lissauer_araa}. The Roman Galactic Exoplanet Survey (RGES) was selected in the 2010 decadal survey \citep{mpf,WFIRST_AFTA} to utilize this uniqueness and measure exoplanet demographics beyond the Kepler mission.'s reach. The aim of this survey is to detect low mass planets beyond the snowline of their hosts and measure the masses of the hosts and the planets. The high resolution capability of the Roman telescope will enable it to do mass measurement of both the hosts and the planets.  Currently,  as part of a precursor study for RGES,  we are developing the mass measurement methods for the exoplanets that are detected from the ground.   

OGLE-2012-BLG-0563Lb was first discovered by OGLE collaboration on May 1, 2012 using the 1.3 m Warsaw telescope.  Later, on May 18, 2012,  the MOA collaboration noticed this event as well using the 1.8 m MOA II telescope.  On May 20,  the $\mu$FUN collaboration circulated an alert that the event was peaking at a high magnification, which means that the source star was approaching very close to the lens star on the sky plane. In such a case, there is a high probability that the light curve will show anomalous features around the peak if the lens star hosts a planet.  Both $\mu$FUN  and ROBONET groups started to take frequent observations of this event with telescopes from Chile,  New Zealand and Australia.  On May 21,  MOA reported the detection of an anomaly as a result of its high cadence observations. The event peaked at around May 21 with the maximum magnification of more than 600, around which the event was well covered by several telescopes in New Zealand and Australia. As a result, a quick look at the light curve clearly showed an anomalous asymmetric feature around the peak, meaning that the lens star hosted a companion. The discovery of this planet was led by \citet{fukui15}. 

The event was also observed with the 8.2m high resolution Subaru telescope with AO  (Adaptive Optics) instrument Infrared Camera and Spectrograph (IRCS, Kobayashi et al. 2000) on July 28, 2012.  Note that this was observed only 2 months after the peak of the event.  \citet{fukui15} measured the excess flux on top of the source and published the mass of the host and the planet.  \citet{fukui15} reported that it is a Saturn mass planet orbiting a 0.34 $\msun$star at 1.3 kpc distance.  In the current paper,  we observed the same event with Keck using NIRC2 AO instrument on July,  2018. We report a different mass measurement than the \citet{fukui15} paper.  In our NIRC2 Keck observation, we find that the lens is partially separated from the source, which makes the PSF elongated.  We fit two stars in elongated PSF to detect the lens and source separately - this method is known as image elongation method.  We measured the lens-source relative proper motion and hence,  the measured masses. We found OGLE-2012-BLG-0563Lb is a Jupiter planet orbiting around 0.81 $\msun$ mass host star at a distance of $5.49 \pm 0.56$ kpc.  This work is important since this will be one of the primary methods of measuring the mass of the host stars and the planets with the      
 {\it Nancy Grace Roman Space Telescope} (formerly WFIRST) mission,
which is NASA's next astrophysics flagship mission, to follow the {\it James Webb Space Telescope} (JWST). 

The {\it Nancy Grace Roman Space Telescope} ( hereafter, {\it Roman} telescope) \citep{WFIRST_AFTA} includes the RGES \citep{penny19} , 
that will detect and measure the mass of hundreds of exoplanets in orbits 
extending from the habitable zone to infinity (i.e.\ unbound planets). The microlens exoplanets discovered by
{\it Roman} will not require follow-up observations since the {\it Roman} observations will also have 
high enough angular resolution to detect the lens (and planetary host) stars \citep{bennett07}. The RGES will observe the galactic bulge in six to eight epochs across the 5 year time-span. Hence the maximum time separation between two epochs will be about $\sim$ 4.5 years. This is because, in spite of the prime mission being 5 years long, the time difference between the first and the last galactic bulge season will be about 4.5 years. For example, a planetary event detected in the first epoch, will be going through lens-source separation in the final epoch.  For an average of 5 mas/yr relative lens-source proper motion,  the separation between lens and source in 4.5 years would be around 23 mas.  The lens separation measured in this paper is $23.28 \pm 0.48 $ mas . Roman will measure masses of a lot of events like this.   

 This event is observed as part of our  
NASA Keck Key Strategic Mission
Support (KSMS) program whose main purpose is to refine the mass measurement method, and 
optimize the {\it Roman} exoplanet microlensing survey observing program. We are conducting a systematic exoplanet microlensing event high angular resolution follow-up program
to detect and determine the masses of the exoplanet host stars with our NASA  Keck KSMS program \citep{bennett_KSMS}. This is supplemented by HST observations \citep{aparna19hst}
for host stars that are most likely to be detected
\citep{bennett06}. This program has already revealed a number of microlens exoplanet host stars that are 
resolved from the source stars \citep{van20,bennett20,blackman21,terry21}, and others that are still blended
with their source stars, but show a significant elongation due to a lens-source separation
somewhat smaller than the size of the point spread function (PSF) \citep{bennett07,ogle169,aparna18}. This program has also recently provided mass upper limits of a Jovian planet orbiting a White Dwarf host. Another purpose of our NASA KSMS program is to confirm measurements of the small lens-source separation with undersampled pixels ( eg. \citet{bennett20}.  The event reported in this paper is also observed by HST ({\it Hubble Space Telescope}) which will be reported in companion paper Bennett et al. 2026.  Note that, since the lens is likely to be brighter in Infrared passbands, we analyzed Keck $K$ data first.  This paper reports the independent analysis of Keck data.

The paper is organized as follows: 
Section \ref{sec-Followup} describes the details of our high resolution follow up observations and the reduction of the AO images.  After that,  section \ref{sec-Keck} demonstrated the Keck narrow camera image analysis details.In the next section, we describe the process of identifying the lens and source system from the two stars.   In section \ref{sec-murel} we show the process of converting heliocentric to geocentric proper motion.  
Finally in sections \ref{sec-lens} we show the results of lens properties and \ref{sec-discussion}, 
we discuss the exoplanet system properties and the implications of its mass and distance measurement.

\section{Follow up observations}
\label{sec-Followup}

\subsection{Wide Camera}
The event was observed  with the Keck Adaptive Optics (AO) NIRC2 \citep{keckAO}
instrument during the  late night of May 23, 2018.  Five dithered exposures, 
each of 30 seconds, were taken in the $K_S$ short passband ( hereafter, $K$ band) with the wide camera. Each wide camera image covers a 1024 $\times$ 1024 square 
pixel area, and each pixel size is about $39.686 \times 39.686\,$mas. These images were flat field and dark current 
corrected using standard methods, and then stacked using the SWarp Astrometics package \citep{SWarp}. 
The details of our methods are described in \citet{batista2014}. We used aperture photometry method on these 
wide images with SExtractor software \citep{sextractor}. These wide images were used to detect 
and match 22 bright isolated stars to the VVV catalog  \citep{vvv} for calibration purposes. The 
average FWHM of this wide camera stack image is 90 mas.  Note that, in the 
2018 wide camera images, the lens and source were not resolved. 
As a result, we need NIRC2 narrow camera images to resolve and to identify the lens system.   
\subsection{Narrow Camera}

This event was also observed with the Keck NIRC2 narrow camera on May 23, 2018 in the $K$-band using laser guide star adaptive optics (LGSAO).  The 
main purpose of these images is to resolve the lens host star from the source star.  Forty dithered observations were taken with
60 seconds exposures. The images were taken with a small dither of 0.7'' at a position angle (P.A.) of 0$^{\circ}$ with each frame consisting of three co-added 20 seconds integration. 
The average FWHM of these images is $\approx$63 mas.  However,  about 10 initial images were lost due to a guiding system problem.  And a few more were lost due to unexpected cloud coverage.  Later we tried to observe this event again on May 25 and May 26, 2018 . We took 23 exposures on May 25 and 10 exposures on May 26. However, due to cloud coverage we could not use Laser Guide Star AO  system and had to rely on only Natural Guide Star AO system. These images demonstrated a much worse seeing,; the  FWHM varying between  120-180 mas. For final reduction, only the best Keck images, that is, the Keck images observed with LGSAO from May 23, 2018 were used.  For the reduction of these images, we used five dark images of 60 seconds exposure time and 10 $K$-band dome flats 
taken with narrow camera on the same day as the science images. There were five dome flat  images with 
the lamp on and five more images with the lamp off, each with 60 seconds exposure time.  Also at the end of 
the night, we took 8 sky images using a clear patch of sky at a (RA, Dec) of (20:29:57.71, -28:59:30.01) 
with an exposure time  of 30 seconds each. All these images were taken with the $K$ band filter. These images were used to flat field, bias subtract and 
remove bad pixels and cosmic rays from the eleven raw science images. The strehl ratio of these clean images varied
over the range 0.31-0.43. Finally 
these clean raw images were distortion corrected, differential atmospheric refraction corrected and stacked into one image \citep{jlu_thesis, KAI}. We used the stacked image for the final photometry and
astrometry analysis.


There are 1024$\times$1024 pixels in each narrow camera image with each pixel subtending
9.942 mas \citep{refraction, distortion} on each side. Since the small field of these narrow images covers only a few bright stars, 
it is difficult to directly photometrically calibrate them to VVV. On the contrary, wide images cover 1024$\times$1024 pixels with each pixel subtending 39.686 mas on each side. So we use the wide camera $K$  stack image that 
was already calibrated to VVV (see above section 3.1) to calibrate the narrow camera images. The wide and narrow images were taken in same filter, hence we did not need to do any filter conversion.
This gives us the brightness calibration between the stacked narrow camera image to VVV image. The 
photometry used for the narrow camera image calibration is from DAOPHOT analysis (section \ref{sec-Keck}). 

\section{Keck Narrow Camera Image Analysis} 
\label{sec-Keck}
In this section, we use DAOPHOT \citep{Daophot} to construct a proper empirical PSF model to identify the two stars (the lens and the source) in the narrow stack images. We started our analysis with 
the 2018 Keck narrow camera images. We used the same method as \citet{aparna18} to build PSF models for the narrow camera stack images. We built these 
PSF models in two stages. In the first 
stage, we ran the FIND and PHOT commands of DAOPHOT to find all the possible stars in the 
image. In second stage, we used the PICK command to build a list of bright ($K < 18.5$) isolated stars that can be 
used to construct our empirical PSF model. Our target object was excluded from this list of PSF stars because it
is expected to consist of two stars that are not in the same position. From this list, we selected seventeen stars close 
to the target that had sufficient brightness, and we built our PSF model from these stars. 

\begin{figure}
\epsscale{1.0}
\plotone{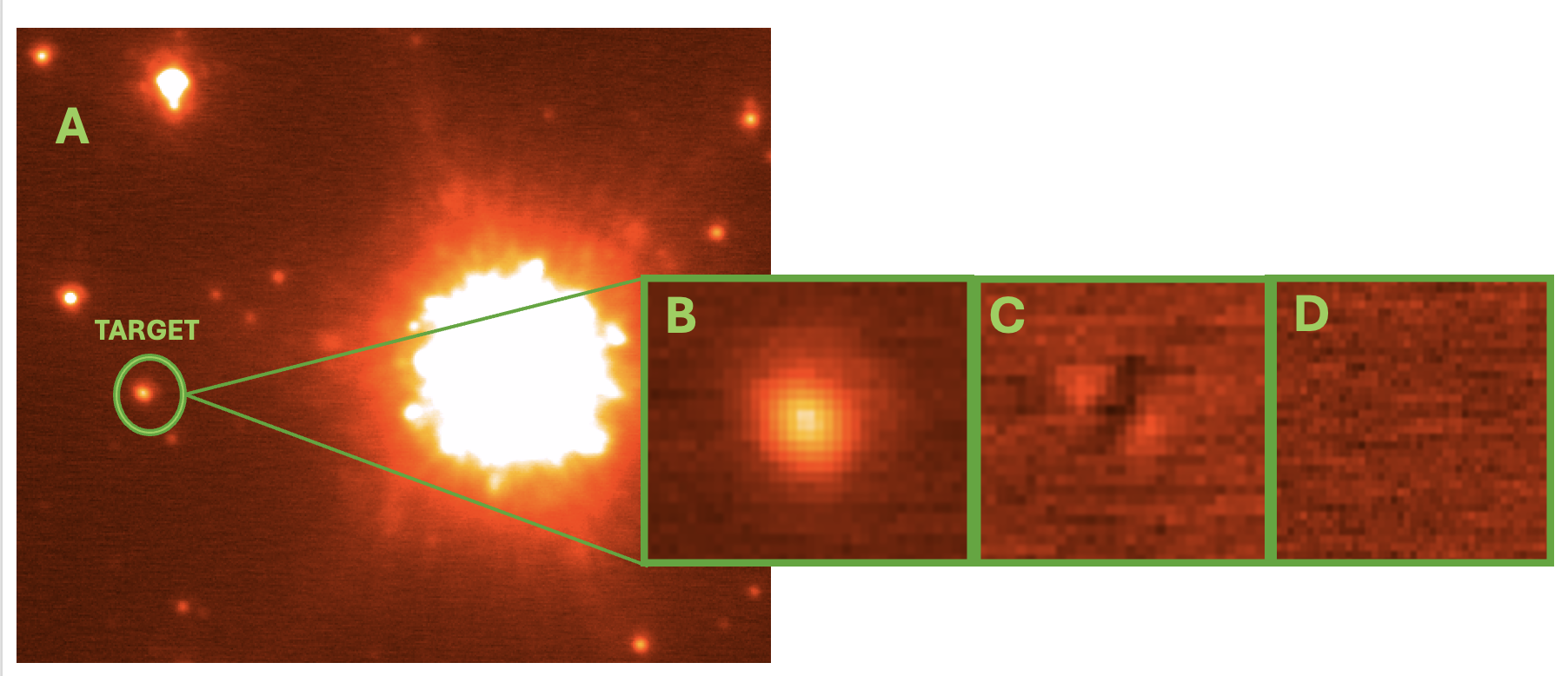}
\caption{{\it Panel A}: A 3" section of the stack image of 20 Keck $K$-band images, taken with the narrow camera,
with the green circle indicating the target.
{\it Panel B}: A close-up of the target object. The source and lens 
positions are obtained from the best fit dual star PSF model. {\it Panel C}: The residual image after subtracting 
the best fit single star PSF model. The residual shows a clear indication of the presence of an additional star since the residual shows a dipole. {\it Panel D}: The residual image after 
subtracting the best fit dual star PSF model. This shows a smooth residual, confirming the second star at 
 separation of $\sim$23 mas. (C) and (D) use the same photometry scale.}
\label{fig-keck1}
\end{figure}

 Once we have built the PSF model, we fit all the stars with this model.  
This step produced the single star residual fit for the target that is shown in Figure \ref{fig-keck1}C. Inspection of this residual image from the single star fit 
indicates that there is an additional star near the target object. So, we tried fitting the region of the target object with a dual star model. The dual star fits produced a smoother residual than the single star fit, as shown in Figure \ref{fig-keck1}D. The results of these dual star fits are given in Tables \ref{tab-Keck_phot} and \ref{tab-murel}. Both the single star and dual star fits were done using the Newton-Raphson method of standard DAOPHOT. 
As we discuss in the next Section, we identify the star in the NE as the lens and the one in the SW as the source  for the
OGLE-2012-BLG-0563 microlensing event.


We have calculated the error bars using 
the Jackknife method \citep{quenouille1949,quenouille1956, tukey1958, tierney} following \citet{aparna21}. This method is 
able to measure the variance due to the 
PSF variations in the individual images. There were a total of 20 clean Keck images. In this method, we built 20 different stacks of 19 images, by removing one of the 20 images each time. 
Then, these 20 images are analyzed with DAOPHOT to build empirical PSFs for each stack image.   Next, the target in each image is fit with the dual star PSF models. Once we have 20 sets of positions and flux of lens and source, the sample mean and the standard deviation of the parameters are used for the statistics. The standard error of a parameter $x$ in Jackknife is given by: 
\begin{equation}
\label{eq-jackknife}
SE(x) = \sqrt{\frac{N-1}{N} \sum (x_{i} - \bar{x})^{2}} \ .
\end{equation}  
The $x_{i}$ represents the value of the parameter measured in each of the combined image and $\bar{x}$ represents the mean of the parameter $x$ from all the N stacked images. This Equation \ref{eq-jackknife} is the same 
formula as the sample mean error, except that it is multiplied by $\sqrt{N-1}$. These error bars are reported in Table \ref{tab-Keck_phot}. The error bars in Table \ref{tab-murel} 
are based on the error bars in Table \ref{tab-Keck_phot} because the Heliocentric lens-source relative proper motion, 
$\mu_{\rm rel,H}$, is proportional to the lens-source separation.

\begin{deluxetable}{ccccc}
\tablecaption{Source and Lens Photometry and Astrometry from Keck data\label{tab-Keck_phot}}
\tablewidth{0pt}
\tablehead{ \multicolumn{2}{c}{Calibrated Magnitudes}&\multicolumn{3}{c}{Lens- Source Separation(mas)} \\
Source &Lens& East& North& total}
\startdata
$17.861 \pm 0.18$ & $17.79 \pm 0.19$ & $20.04 \pm 0.18$ & $11.85 \pm 0.45$ & $23.28 \pm 0.48$ \\
\enddata
\tablecomments{The separation was measured 6 years after the peak of the event in the 
$K$ band. The follow-up observation was taken on May 23, 2018 and the peak of the microlensing light curve event was on May 21, 2012.}   
\end{deluxetable}

\section{Identifying the Lens and the Source}
  In Keck analysis, the two stars have similar brightness : $17.86 \pm 0.18$ and  $17.79 \pm 0.19$. Hence it is difficult to know which one is source from the $K$ band magnitudes.  To determine the correct source,  we did two light curve fitting with the constraints from the Keck image analysis.  In the first case, case A) is where the star in North- East (NE) is considered the lens star and the star in South-West (SW) considered to be the source star.  In the second case scenario, case B,  the lens and the source stars are switched - source is in NE and lens is in SW.   
  
  \subsection{Re-reduction of the Light Curve Data}
  
    We did this light curve analysis using  re-reduced MOA photometry using \citet{bond17}.  This re-reduced photometry was obtained by selecting MOA observations from mid 2011 to the end of 2016. Difference
imaging was used to derive the photometry, with each of the MOA passbands treated separately with their own reference images.  Microlensing events are observed in crowded fields and
their centroids on the images are often blended with neighbouring stars. It is important that these centroids are measured carefully because the nearest resolved star on the image may not necessarily be the source star for the microlensing event. The MOA difference images are measured using
a reference image analytical PSF model similar to the form of
the Dophot photometry code of Schechter et al. (1993). This
PSF is then used to measure a given difference image.  The
model PSF is optimized by finding the centroid and shape
parameters that give the best photometry on a set of difference images where the flux of the source star is significantly
magnified.
The baseline photometry, in $R$ and $V$ bands separately,
was examined during the off-event years (ie 2011–2015) for
any correlations that may be present due to variations in the
seeing and the effects of differential refraction (parameterized by the hour angle). We find some small effects present
and so we “detrend” the data by modelling the baseline as a
function of seeing and the hour angle.  These respective $R$ and $V$ seeing models were used as corrections to
photometry which were subtracted off all the data including
those where the source is magnified.   This method was not available or implemented in \citet{fukui15} analysis and an improvement in our current analysis. 

\subsection{Image Constrained Light Curve Fitting}   
   
    We fitted the light curve data with lens flux and heliocentric relative proper motion direction constraints from the Keck analysis.  Our detection of the exoplanet host star in the high angular
resolution follow-up observations can be used to determine the
masses and distance of the host star and planet along with the
parameters  from the microlensing light-curve models. One common approach (D. P. Bennett et al. 2015) is
to generate a collection of model parameters that are consistent
with the light curve using MCMC calculations, and then to
exclude those models that are inconsistent with the high
angular resolution follow-up observation results.
In cases where the modeling includes the microlensing
parallax effect ($\pi_E$), the fraction of light-curve models in the
Markov Chain that are consistent with the high-resolution
follow-up data can be very small. This is largely because the
light-curve measurement of the source-lens relative proper
motion constrains the direction of the parallax vector to be parallel to
the relative proper motion. Although parallax should be present in
every microlensing event observed by ground-based telescopes, it is often not included in light-curve modeling analyses
since its parameters cannot be measured very precisely. This
approach can be problematic, however, because even when $\pi_E$
parameters cannot be measured precisely, they can still
influence some of the other model parameters. Therefore, it is
important to include parallax values that are consistent with the high
angular resolution follow-up data when exploring light curve models.
The vast majority of models from an MCMC run (including
$\pi_E$ models) are unlikely to be consistent with the follow-up
data, which means that very long and time consuming MCMC
runs will be needed to get reasonable sampling of the lightcurve models consistent with the data. To address this problem,
we have modified our fitting code, eesunhong16 (D. P. Bennett
$\&$ S. H. Rhie 1996; D. P. Bennett 2010), to include the
constraints from the position, direction and brightness of the lens star
identified in the Keck AO images.  In this event, for case A and case B,  the parallax directions will be opposite.  This constraint from image analysis combined with the light curve modeling will determine which case is more likely. 
    
     The heliocentric proper motion measured in Keck image is calculated from Table 1 and reported in Table 2.  This heliocentric proper motion direction is opposite between case A and case B,  which would yield different values and directions for geocentric relative proper motion $\bm{\mu}_{\rm rel,G}$ and hence for geocentric microlensing parallax .  As \citet{gould14} shows:
     \begin{equation}
\bm{\mu}_{\rm rel,H} = \bm{\mu}_{\rm rel,G} + \frac{{\bm v}_{\oplus} \pi_{\rm rel}}{\rm AU} =\frac{\theta_E \bm{\pi}_{\rm E,geo} }{t_E \pi_E} +   {\bm v}_{\oplus} \pi_E \theta_E   \ ,
  \end{equation}
    From above equation,  we know ${\bm{\pi}_{\rm E,geo}}$ is parallel to ${\bm{\mu}_{\rm rel,G}}$ and  ${\bm{\mu}_{\rm rel,G}}$ is different for case A and case B as the ${\bm{\mu}_{\rm rel,H}}$ has opposite direction for case A and case B.  Hence parallax has different directions for case A and case B.  We did image constrained light curve modeling with a 2L1S (two lens one source) model,  with constraints on relative proper motions and hence constraints on parallax directions.  These different directions of parallax made a significant impact on the modeling.  
    
 We fitted several light curve models to the re-reduced MOA data combined with FTS and MOA B $\&$ C data.  The results for the unconstrained, magnitude-constrained, and constrained Case A and Case B models (where both magnitude and relative proper motion directions are constrained) are presented in Table 3.  The $\Delta\chi^{2}$ between the unconstrained model and the magnitude constrained model was only $\sim$2. Notably both case A and case B solutions had higher $\Delta\chi^{2}$ than the  model with only $K_L$ magnitude constraint by 45.94 and 36.35,  respectively.  However the $\Delta\chi^{2}$ should have increased by only 2 for the two directions of ${\mu}_{\rm rel,H}$ .  This discrepancy indicated an underlying issue with the light curve modeling.  

To investigate the source of this difference, we examined the light curve data more closely and found that FTS data and MOA B$\&$C $I$ band data did not fit the light curve model well, even without constraints.  This is shown in Figure 4.  With the lens flux and relative proper motion direction constraints,  they were contributing to higher $\Delta\chi^{2}$ - about $\sim$26 for case A and $\sim$17 for case B.  However,  for OGLE data the $\chi^{2}$ actually improved with the constraints from high resolution observations in case A by $\sim$5 and got worse for case B by $\sim$14.  We would point out that the parameters that were most affected are : the source radius crossing time and the parallax.  However,  FTS and MOA B$\&$C $I$ band data are short duration data which could not significantly constrain the parallax.  However, OGLE and MOA data were able to constrain the parallax better because of their longer baseline.  In addition,  the OGLE photometry is observed to be the most reliable in the analysis of other events.  All of this signals photometric systematics in FTS data and MOA B$\&$C $I$ band data.  Furthermore,  the time period covered by the FTS data is also covered by a number of other data
sets including MOA and the Perth Exoplanet Survey Telescope (PEST), and these data sets do
not favor the deviations from the light curve models seen by FTS.  Because,  MOA B $\&$C also has shown systematics in previous analysis of other events,  we removed those two data sets and reanalyzed the light curve models for case A and case B.  

We notice the $t_*$ increased about 4 times from the unconstrained fit to the constrained case A in Table 3.  Note that both the constrained fits had higher $t_*$ than the unconstrained fit.  This change affected the Einstein ring radius and consequently,  the relative proper motion as well.  The geocentric relative proper motion is highest for the unconstrained case where the $t_*$ is smallest among the four fits and the relative proper motion is lowest for case A where $t_*$ is biggest. 

We fitted several light curve models to the data without FTS and MOA B $\&$ C data.  These fits are reported in Table 4.  There are three fits shown - unconstrained and case A and case B.  As shown, case A has a  $\Delta\chi^{2}$ $\sim$7  compared to the unconstrained fit and case B has a $\Delta\chi^{2}$ $\sim$18 compared to the unconstrained fit.  But in both cases there are only three additional constraints added: magnitude and two directions of the lens-source proper motion.  Both OGLE and MOA data, which are the usually reliable data-sets,  seem to indicate worse  $\Delta\chi^{2}$  for case B.  This indicated that case A, that is,  the case where the lens is the North-East candidate is the correct solution.  We fitted the light curve data without FTS and MOA B $\&$ C data and reported our final light curve modeling parameters in Table 4.  We also added the final light curve modeling parameters from \citet{fukui15} for comparison in Table 5. This difference is mainly due to the re-reduced MOA photometry, removal of FTS and  MOA B$\&$C data and constrained light curve modeling with relative lens-source proper motion.  The light curve graph does not show significant difference from
\citet{fukui15} paper, { \bf however the new constrained fit is shown in Figure 3}.  The FTS and MOA B$\&$C data were favoring the light curve model with higher $t_E$ value.  But our final best fit light curve model has a smaller $t_E$ value which also means a higher mass ratio value.  The smaller $t_E$ value is accompanied by smaller $\theta_E$ and smaller $\mu_{\rm rel,G}$ value.  It is to be noted that the case B showed a $\mu_{\rm rel,G}$ value closer to the unconstrained fit from the re-reduced MOA data and removing FTS,  MOA B$\&$C data.  However MOA and OGLE data favored case A.  Hence if we analyzed the light curve without any $\mu_{\rm rel,G}$ constraint from image,  we might have mistaken case B as the right solution. 
    
  \subsection{Consistent Measurements with HST}
  
  This event was also observed in 2018 May 23 with HST in $I$ and $V$ bands.  There were 16 images observed in $I$ band (F814W) and 16 images in $V$ band (F555W).  The exposure times of these images are 85 seconds for $I$ band and 92 seconds for $V$ band.  From the analysis of these HST images, it was found that the star in the south-west in $I$ band is consistent with the predicted source magnitude from the light curve modeling,  and the star in North-East is fainter and redder than the star in South-West.  This indicated the star in south-west is lthe source and star in North east is likely to be the lens and confirmed the case A.  This is described in our companion paper Bennett et al. 2026.  Note that the lens was very faint and barely detected in $V$ band.  The separations obtained for both HST $I$ and $V$ are consistent with Keck results shown in Table 1. The $\mathbf{\mu}_{\rm rel,H}$ for $I$ band is ($\mathbf{\mu}_{\rm rel,H, N} = 2.56 \pm0.21 , \mathbf{\mu}_{\rm rel,H, E} = 3.51\pm 0.19$).  Similarly for $V$ band,  the measured relative proper motion is:  ($\mathbf{\mu}_{\rm rel,H, N} = 2.26 \pm0.28 , \mathbf{\mu}_{\rm rel,H, E} = 3.45\pm 0.25$). These values are consistent with the relative proper motions measured with Keck, as shown in Table \ref{tab-murel}: ($\mathbf{\mu}_{\rm rel,H, N} = 1.97 \pm0.23 , \mathbf{\mu}_{\rm rel,H, E} = 3.34\pm 0.13$).  The Figure 2.  shows the comparison between HST and Keck analysis.  Since we got consistent results in all 3 passbands,  we can say confidently that these two stars are lens and source system and not a nearby non-relevant additional star.  Note that even though the lens was very faint in $V$ band,  since the lens was detected in both $K$ and $I$ bands,  the lens-source separation was definitive and was used as a constraint to detect the lens in $V$ band.  The lens is brightest in $K$ band and without the Keck image analysis,  only $I$ band detection would not necessarily guarantee the confirmation of lens, specially because the measured separation being very different that the original discovery paper \citet{fukui15}.  
  
  \subsection{Probability of a Nearby star or a Companion to the Source or Lens}
     We did an analysis using the method of \citet{highres} to find out the probability of the detected lens star to be an unrelated star in the bulge or a companion to the source.  Using genulens \citep{genulens} we generated stars around the event position, $(l,b) = (3.31, -3.25)$,  over a 0.05 deg $\times$ 0.05 deg field.  We selected stars with brightness same as lens star.  Using a Poisson probability for finding at least one such star within 23.28 mas of the source star,  we get  $0.0025\%$.  Hence,  the probability of a nearby star mimicking to be a lens star is $0.0025\%$.  For the source-companion case,  we generated source-like stars with $K = 17.861 \pm 0.18$ and used genstars from genulens with BINARY=1.  We then counted companions with $K$ magnitude of  $17.79 \pm 0.19$ and projected separation smaller than 23.28 mas. This gives the probability of source companion is $0.01\%$. We also found that the probability of any of this star of $K$ magnitude of  $17.79 \pm 0.19$ being a companion to lens is $<0.1\%$. 
\begin{figure}
\epsscale{1.0}
\plotone{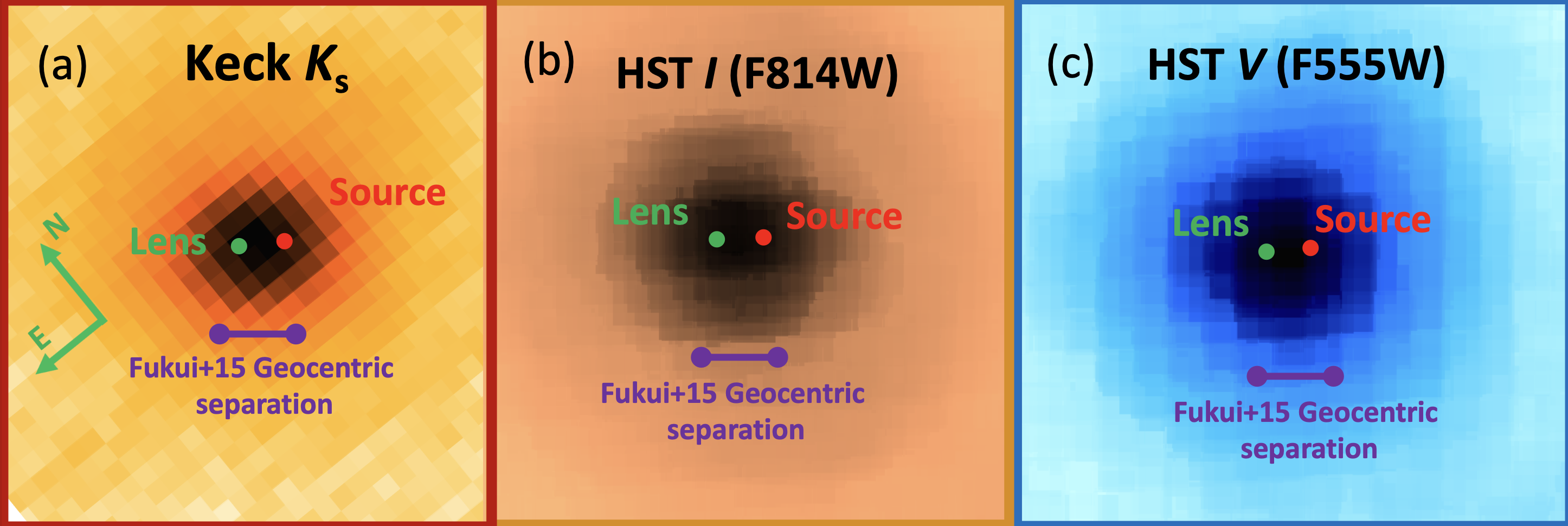}
\caption{The left shows the Keck K band image and same as Fig. 1 B. The middle image is HST $I$ band stack image and the right one is HST $V$ band stack image of the same target (Bennett et al. 2026). The lens and source are plotted in three passbands. The heliocentric lens-source separations are consistent with each other,  but smaller than the predicted geocentric separation from \citet{fukui15}. }
\label{fig-keck}
\end{figure}

\begin{figure}
\epsscale{1.0}
\plotone{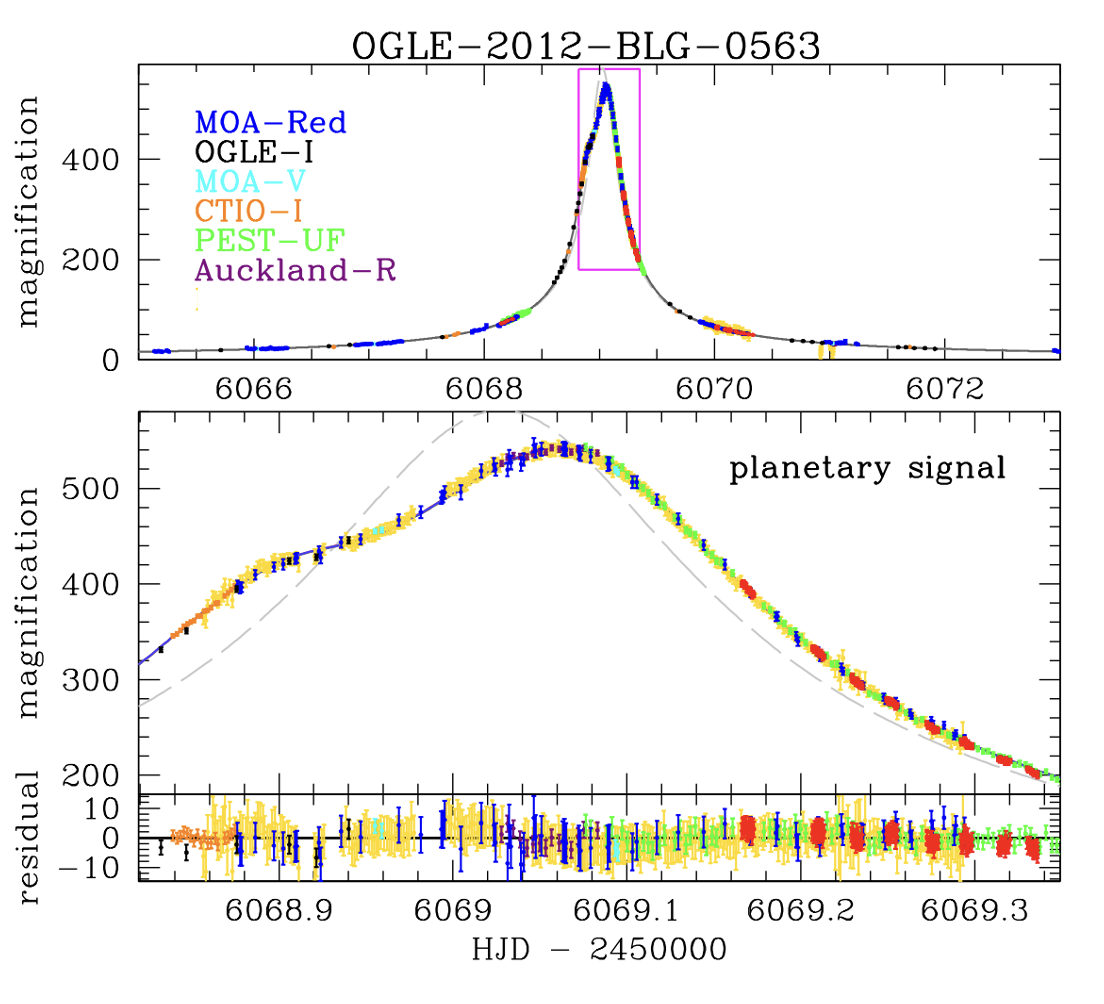}
\caption{The final image constrained fit light curve model of OGLE-2012-BLG-0563 with re-reduced MOA photometry is shown.  The lower box shows the planetary signal. The dashed line represents single lens light curve.}
\end{figure}

\begin{figure}
\epsscale{1.0}
\plotone{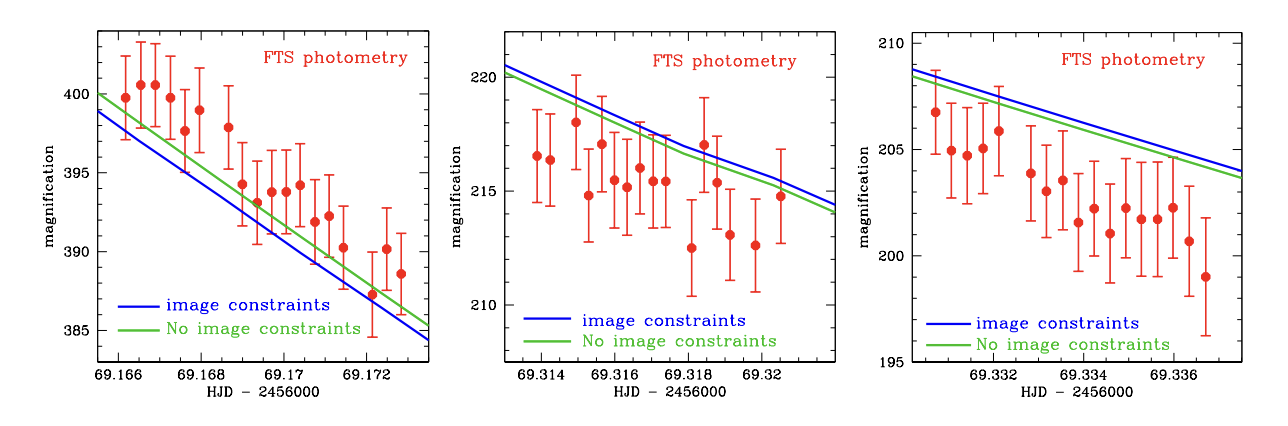}
\caption{A close-up comparison of the FTS data from the first and last two 10 minute observing
intervals on the night of the light curve peak. This data is compared to best fit models with and
without the inclusion of constraints from the high angular resolution images. The model light curves
obtained with the constraints are shown in blue, and the light curves without these constraints are
shown in green. The difference between the model curves is much smaller than the offset between
the data and the models. It shows that the FTS data is not a good fit to either model.}
\label{fig-fts}
\end{figure}
I
l

\begin{deluxetable}{cccc}
\tablecaption{Measured Lens-Source Relative Proper Motion\label{tab-murel}}
\tablewidth{0pt}
\tablehead{
\multicolumn{3}{c}{$\mathbf{\mu}_{\rm rel,H}$(mas/yr)}&\multicolumn{1}{c}{$\mathbf{\mu}_{\rm rel,G}$(mas/yr)}\\
 $\bm{\mu}_{\rm rel,H,E}$&$\bm{\mu}_{\rm rel,H,N}$ & $|\mathbf{\mu}_{\rm rel,H}|$ &}
\startdata
   $3.34\pm 0.13$ & $1.97 \pm 0.23$ & $3.88 \pm 0.26$ &  $3.69 \pm 0.04$  \\
\enddata
\tablecomments{$\mu_{\rm rel, G}$ from Keck in this Table is obtained from the measurement of $\mu_{\rm rel, H}$ using the Equation \ref{eq-mu_helio} and constraining the light curve using image constrained light curve modeling method with re-reduced MOA photometry and removing systematics from FTS and B$\&$C data.  }
\end{deluxetable}

\begin{deluxetable}{ccccc}
\tablecaption{Light Curve Models with re-reduced MOA-Red and FTS and B$ \&$C data and Different Constraints\label{tab-lckeck}}
\tablewidth{0pt}
\tablehead{Parameters&Unconstrained&$K_L$ constrained&Case A&Case B}
\startdata
$t_E$ (days)&62.058&62.157&63.979&64.081 \\
$t_0$ (HJD')&6069.0290&6069.0291&6069.0275&6069.0276\\
$u_{min}$&-0.001754&0.001755&0.0017192&$-0.0017119$\\
s&0.40159&0.40559&0.41513&0.41118\\
$\alpha$ (rad)&0.49856&0.49769&0.49397&$-0.49705$\\
$q$($\times 10^{-3}$)&1.4891&1.4583&1.3847&1.3917\\
$t_*$ (days)&0.01951&0.02488&0.04004&0.03516\\
$\pi_{E,N}$&0.59927&0.60754&0.05361&$-0.06788$\\
$\pi_{E,E}$&0.11011&0.11019&0.08207&$-0.12935$\\ 
$\theta_E$ (mas)&1.532&1.201&0.673&0.817\\
$D_L$ (kpc)&$-$&1.15&5.58&4.14\\
$\mu_{rel,G}$(mas/yr)&9.0101&7.0603&3.8424&4.656\\
$\chi^2_{\rm Ogle\  I}$&670.04&667.8&662.199&681.31\\
$\chi^2_{\rm fts\  I}$&185.52&187.551&213.516&204.316\\
$\chi^2_{{\rm MOA}\  R}$&164.804&164.231&165.391&170.354\\
$\chi^2_{\rm B\&C}$&528.29&529.838&542.155&533.883\\
Total $\chi^2$&2064.65&2066.18&2112.12&2102.53\\
\enddata
\tablecomments{Case A and Case B are both lens magnitude and heliocentric relative proper motion constrained fit.  The direction of the relative proper motion constrains the parallax directions.  Case A corresponds to the lens being the NE star and case B corresponds to the source being the NE star.  Case A and Case B both show much higher $\chi^2$ that expected which seem to come from systematics of FTS and MOA B$\&$C data. After removing those two data,  Case A turns out to be the correct solution. }
\end{deluxetable}

\begin{deluxetable}{ccccc}
\tablecaption{Light Curve Models with re-reduced MOA and no FTS + MOA B$\&$C data}
\tablewidth{0pt}
\tablehead{Parameters&Unconstrained&Case A& Case B& Final Fit MCMC Averages}
\startdata
$t_E$ (days)&62.189&65.72&63.004&63.8 $\pm$2.1\\
$t_0$ (HJD')&6069.0284&6069.0280&6069.0282&6068.96$\pm$0.74\\
$u_{min}$&-0.0017650&0.0016786&-0.0017486&-0.00115$\pm$0.00062\\
s&0.41438&0.42669&0.42310&0.42304$\pm$0.00851\\
$\alpha$ (rad)&-0.49644&0.49226&-0.49336&0.49281$\pm$0.00272\\
$q$($\times 10^{-3}$)&1.3962&1.2615&1.3207&1.331$\pm$0.085\\
$\pi_{E,N}$&-0.34704&0.05661&-0.05813&0.0583$\pm$0.0060\\
$\pi_{E,E}$&0.12073&0.0859&-0.10507&0.0801$\pm$0.0046\\
$\mu_{rel,G}$(mas/yr)&5.1805&3.6699&4.2933&3.694$\pm$0.038\\
$D_L$ (kpc)&$-$&5.1723&4.8303&5.49$\pm$0.56\\
$\chi^2_{\rm Ogle\  I}$&659.784&660.586&672.573&$-$\\
$\chi^2_{{\rm MOA}\  R}$&154.34&159.188&163.367&$-$\\
Total $\chi^2$&1338.21&1345.29&1355.91&$-$\\
\enddata
\tablecomments{Case A corresponds to the lens being the NE star and case B corresponds to the source being the NE star.  Both for Case A and B, the light curve models are constrained by the lens magnitude and the direction of the relative proper motion. Hence they have 3 additional constraints.  As seen Case A has lower $\chi^2$ indicating the correct solution of lens and source identification. }
\end{deluxetable}

\begin{deluxetable}{ccc}
\tablecaption{Final Light Curve Model Comparison from 2026 and \citet{fukui15} \label{tab-lc}}
\tablewidth{0pt}
\tablehead{Parameters&2026 Analysis& 2015 Analysis}
\startdata
 $t_E$(days)&$63.8\pm2.1$&$77.5\pm2.2$\\
 $t_0$(HJD')&$6068.96 \pm 0.7$ &$6069.03 \pm 0.01$ \\
$\theta_E$(mas)&$0.65\pm 0.02$&$1.36\pm0.13$ \\
$q$($\times 10^{-3}$)&1.331$\pm$0.085&$1.09\pm0.04$ \\
$\mu_{rel,G}$(mas/yr)&$3.69\pm0.04$&$6.4\pm0.5$\\
\enddata
\tablecomments{Current analysis is done with rereduced MOA data, OGLE data, $\mu$FUN and Robonet data; FTS and MOA B$\&$C data are removed.  \citet{fukui15} analysis included FTS and MOA B$\&$C data,  but MOA $R$ data was not treated with detrending.  The final light curve results differ due to data quality.}
\end{deluxetable}

\section{Determination of Geocentric Relative Lens-Source Proper Motion}
\label{sec-murel}

 We determined the relative proper motion between the lens and the source star.  In this subsection, we report measuring the heliocentric relative proper motion from narrow Keck images and then transform that to geocentric frame.  Then we compare the measured geocentric relative proper motion to the predicted relative lens-source proper motion from the light curve, given in Table \ref{tab-lc}. 
 
The two stars were separated by one-third of the FWHM of the $K$ band image,with the separations given in Table \ref{tab-Keck_phot}.
 The measured separations can be used to compute the lens-source relative proper motion, $\mu_{\rm rel}$, 
 which can be compared with the $\mu_{\rm rel}$ prediction from the light curve.  However, this issue is complicated
by the fact that the lens-source separation measurements determine the relative proper motion in a Heliocentric
frame, $\mu_{\rm rel,H}$, while the light curve measures the relative proper motion in an inertial Geocentric reference
frame, $\mu_{\rm rel,G}$, that moves with the Earth's velocity at the time of the event. The relationship between
these reference frames is given by equation \ref{eq-mu_helio}.

 Our high resolution Keck observations of 2018 were taken $\sim$6 years after the microlensing event magnification peak. Since these images were taken almost exactly 6 years after the microlensing magnification, we do not need to consider the relative position of the Earth with respect to the Sun when determining the heliocentric lens-source relative proper motion
from the measured separations in the Keck images.

At the time of peak magnification, the separation between lens and source was 
$\sim |u_0\theta_E| \sim 0.017\,$mas. 
Hence, by dividing the measured separation by the time interval 
of 6 years, we obtain the heliocentric lens-source 
relative proper motion, $\mubold_{\rm rel,H}$. A comparison of these values from our independent dual star
fits for the $K$ band is shown in Table \ref{tab-murel} with 
error bars are estimated from jackknife method. 

Our light curve models were done in a geocentric reference frame that differs from the heliocentric
frame by the instantaneous velocity of the Earth at the time of peak magnification, because the 
light curve parameters can be determined most precisely in this frame. However, this also means that
the lens-source relative proper motion that we measure with follow-up observations is not in the
same reference frame as the light curve parameters. This is an important issue because, as we
show below (see section \ref{sec-lens}), the measured relative proper motion can be combined with brightness of
the source star determine the mass of the lens system. The relation between the relative
proper motions in the heliocentric and geocentric coordinate systems are given by \citet{dong-moa400}:
\begin{equation}
\bm{\mu}_{\rm rel,H} = \bm{\mu}_{\rm rel,G} + \frac{{\bm v}_{\oplus} \pi_{\rm rel}}{\rm AU}  \ ,
\label{eq-mu_helio}
\end{equation}
where ${\bm v}_{\oplus}$ is the projected velocity of the earth relative to the sun (perpendicular to the 
line-of-sight) at the time of peak magnification. 

The projected velocity for OGLE-2012-BLG-0563 is
${{\bm v}_{\oplus}}_{\rm E, N}$ = (25.107, 1.446) km/sec = (5.29, 0.31) AU/yr at the peak of the microlensing. The relative parallax is 
defined as $\pi_{\rm rel} \equiv  (1/D_L - 1/D_S)$, where $D_L$ and $D_S$ are lens and source distances. Hence the Equation \ref{eq-mu_helio} can written as:
\begin{equation}
\bm{\mu}_{\rm rel,G} = \bm{\mu}_{\rm rel,H} - (5.29, 0.31 )\times (1/D_L - 1/D_S) 
\label{eq-mu}
\end{equation}
Since $\bm{\mu}_{\rm rel,H}$ is already measured in Table \ref{tab-murel}, Equation 3 yields the geocentric relative proper motion, $\bm{\mu}_{\rm rel,G}$ as a function of the lens distance.  Now at each possible lens distance, 
we can use the $\mu_{\rm rel,G}$ value from equation~\ref{eq-mu} to determine the angular 
Einstein radius, $\theta_E = \mu_{\rm rel,G} t_E$. Since we already know the $\theta_E$ value from the light curve models, we can use that here to constrain the lens distance and relative proper motion. Using this method, we determined the
relative proper motion to be $\mu_{\rm rel,G} = 3.69\pm 0.04$ mas/yr.  Note that, the value of relative proper motions depends on the identification of the lens.  For case A, which is the preferred scenario,  $\mu_{\rm rel,G} = 3.69\pm 0.04$ mas/yr.  However, for case B,  the heliocentric proper motion direction is reversed and $\mu_{\rm rel,G} = 4.29 \pm 0.06$ mas/yr.  The relative proper motion for case B is actually closer to the $\mu_{\rm rel,G}$ predicted by the unconstrained light curve model with re-reduced MOA photometry and FTS and B$\&$C data removed.  Even though for case B,  the geocentric relative proper motion is bigger and closer to \citet{fukui15} prediction and the unconstrained fit prediction from Table 4,  case B is not the correct identification since it yields higher $\chi^2$ value than case A by $\sim$11 (See Table 4) and case A is also consistent with HST measurements (Bennett et al. 2026 and section 4.3) .  For case B,  the constraints on parallax yields a higher $\chi^2$ by $\sim$10, hence we know the final  $\mu_{\rm rel,G} = 3.69\pm 0.04$ mas/yr.

\section{Lens Properties}
\label{sec-lens}
Mass measurements of microlensing exoplanets and their hosts can be done solving a minimum of two of the three mass distance relations. 
Measurements of the angular Einstein radius, $\theta_E$, and the microlensing parallax 
amplitude, $\pi_E$, can each provide mass-distance relations \citep{bennett_rev,gaudi_araa},
\begin{equation}
M_L = {c^2\over 4G} \theta_E^2 {D_S D_L\over D_S - D_L} = {c^2\over 4G}{ {{\rm AU}^2}\over{\pi_E}^2}{D_S - D_L\over D_S  D_L}  \ .
\label{eq-m_thetaE}
\end{equation}
 $D_L$ and $D_S$ are distances to the lens and the source respectively. These can be combined to yield the lens mass in an expression with no dependence on the lens or source distance,
\begin{equation}
M_L = {c^2 \theta_E {\rm AU}\over 4G \pi_E} = {\theta_E \over (8.1439\,{\rm mas})\pi_E} \msun \ .
\label{eq-m}
\end{equation}
The angular Einstein radius can be measured for most planetary microlensing events because 
most planetary events have finite source effects that allow the measurement of the source radius crossing time,
$t_*$. The angular Einstein radius is then given by $\theta_E = \theta_* t_E/t_*$, where $t_E$ is the Einstein radius crossing time and
$\theta_*$ is the angular source radius, which can be determined from the
source brightness and color \citep{kervella_dwarf,boyajian14}.  As a result, the measurement of $\pi_E$ for planetary events
usually results in mass measurements. Unfortunately, the orbital motion of the Earth allows $\pi_E$ 
to be determined for only a relatively small subset of events such as the ones that have very long durations
\citep[e.g.][]{gaudi-ogle109,bennett2010}, long duration events with bright source stars \citep[e.g.][]{muraki11},
and events with special lens geometries \citep[e.g.][]{sumi16}. The microlensing parallax program using the
{\it Spitzer} space telescope at $\sim 1\,$AU from Earth has recently expanded the number of 
events with microlensing parallax measurements \citep{udalski_ogle124,street16}, but recent studies
indicates that systematic errors in the {\it Spitzer} photometry can affect
some of the {\it Spitzer} $\pi_E$ measurements \citep{spitz_vs_gal,gould20,dang20}.

In order to obtain good sampling of light curve model parameters that are consistent with our photometric 
constraints and astrometry, we apply the following constraints, along with Galactic model constraints
when summing over our light curve modeling MCMC results to determine the final parameters.
The proper motion vectors of $\mu_{\rm rel,H}$ and the lens magnitudes are constrained according to Table \ref{tab-Keck_phot} and Table  \ref{tab-murel}. 
The $\mu_{\rm rel,H}$ constraints are applied to the Galactic model and the lens magnitude
constraints are applied when combining the MCMC light curve model results with the Galactic model.
The lens magnitude constraints require the use of a mass-luminosity relation. We built
an empirical mass luminosity relation following the method presented in \citet{bennett_moa291}.
This relation is a combination of mass-luminosity relations for different mass ranges. 
For $M_L \geq 0.66\,\msun$, $0.54\,\msun\geq M_L \geq 0.12\,\msun $, and 
$0.10 \,\msun \geq M_L \geq 0.07\,\msun$, we use the relations of \citet{henry93}, \citet{delfosse00},
and \citet{henry99}, respectively. In between these
mass ranges, we linearly interpolate between the two relations used on the
boundaries. That is, we interpolate between the \citet{henry93} and the \citet{delfosse00}
relations for $0.66\,\msun > M_L > 0.54\,\msun$, and we interpolate between the
\citet{delfosse00} and \citet{henry99} relations for $0.12\,\msun > M_L > 0.10\,\msun$. When using
these relations we assume a 0.05 magnitude uncertainty.

For the mass-luminosity relations, we must also consider the foreground extinction.
At a Galactic latitude of $ b = -3.2522^\circ$, most of the dust is likely to be in the foreground of
the lens unless it is very close to us. We quantify this with a relation relating the extinction 
of the foreground of the lens to the extinction in the foreground of the source.
Assuming a dust scale height of $h_{\rm dust} = 0.10\pm 0.02\,$kpc, we have
\begin{equation}
A_{i,L} = {1-e^{-|D_L(\sin b)/h_{\rm dust}|}\over 1-e^{-|D_S (\sin b)/h_{\rm dust}|}} A_{i,S} \ ,
\label{eq-A_L}
\end{equation}
where the index $i$ refers to the passband: $V$, $I$, $H$, or $K$.

These dereddened magnitudes can be used to determine the angular source radius,
$\theta_*$. With the source magnitudes that we have measured, the most precise determination
of $\theta_*$ comes from the $(V-I),I$ relation. We use
\begin{equation}
\log_{10}\left[2\theta_*/(1 {\rm mas})\right] = 0.501414 + 0.419685\,(V-I)_{s0} -0.2\,I_{s0} \ ,
\label{eq-thetaS}
\end{equation}
which comes from the \citet{boyajian14} analysis, but with the color range optimized for the 
needs of microlensing surveys \citep{aparna16}.

\begin{deluxetable}{cccc}
\tablecaption{Comparison of Measurement of Planetary System Parameters from current analysis with constraints from $\mu_{\rm rel, H}$ vs  Fukui et al.  (2015) \label{tab-params-histogram}}
\tablewidth{0pt}
\tablehead{\colhead{parameter}&\colhead{units}&\colhead{Final fit 2026}&\colhead{Final fit Fukui et al.  (2015)}}
\startdata
Geocentric lens-source relative proper motion, $\mu_{\rm rel, G}$&mas/yr&$3.69\pm 0.04$& $6.4^{+0.6}_{-0.5}$\\
Host star mass, $M_{\rm host}$&${\msun}$&$0.81^{+ 0.03}_{-0.04}$ &$0.34^{+0.12}_{-0.20}$\\
Planet mass, $m_p$&$M_{\rm Jup}$& $1.12^{+0.08}_{-0.08}$& $0.39^{+0.14}_{-0.23}$\\
Host star-planet 2D separation, $a_{2D}$&AU&$4.77^{+0.38}_{-0.31}$&$4.3^{+1.5}_{-2.5}$\\
Lens distance, $D_L$&kpc &$5.49^{+ 0.58}_{-0.55}$&$1.3^{+0.6}_{-0.8}$\\
\enddata
\end{deluxetable}

We apply the $K$-band mass-luminosity relations to each of the models in our Markov Chains
using the mass determined by the first expression of equation~\ref{eq-m_thetaE}, using the $\theta_E$
value determined from $\theta_E = \theta_* t_E/t_*$, where $t_E$ and $t_*$ are light curve parameters
given in Table~\ref{tab-Keck_phot}. We can then use the Keck $K$-band measurements of the
lens star brightness from Table~\ref{tab-Keck_phot} to constrain the lens brightness including both the
observational uncertainties listed in Table~\ref{tab-Keck_phot} and the 0.05 mag theoretical uncertainty
that we assume for our empirical $K$-band mass-luminosity relations. We solve the first expression of equation~\ref{eq-m_thetaE} along with the mass-luminosity relations. While solving, we assume the source distance $D_S  = 8.51\pm 1.14$ kpc. We solve this for every MCMC model and we sum over our MCMC results
using the Galactic model employed by \citet{bennett14} as a prior, weighted by the 
microlensing rate and the measured $\mubold_{\rm rel,H}$ values given in Table~\ref{tab-murel}. The results of our final sum over the Markov Chain light curve models are given in 
Table~\ref{tab-params-histogram}. 
This mass measurement assumes uniform planet hosting probability. 
This table gives the mean and the central 68.3\% confidence interval of the parameters. The host mass is measured to be $M_{\rm host} = 0.81\pm 0.04\msun$, a K dwarf star, orbited by a 
Jupiter mass planet,
$M_{\rm Planet} = 1.12\pm 0.09 M_{\rm Jup}$. 

Table \ref{tab-params-histogram} show the comparison between mass measurements from 2018 Keck data and the 2012 Subaru measurements that was presented in Fukui et al. (2015).  During the 2012 Subaru measurements,  only excess flux was detected on top of the source, but relative lens-source proper motion could not be measured.  However, from 2018 Keck data, relative lens-source proper motion was measured and the direction constraint was added to the light curve.  These constraints preferred a different model and hence yielded a different mass measurement. The mass measurement for this event with HST data is similar and consistent: $M_{\rm host} = 0.81\pm 0.04\msun$,  $D_{L} = 5.49^{+0.58}_{-0.55} $ kpc, $M_{\rm Planet}  =1.13^{+0.09}_{-0.08} M_{\rm Jup}$.  More details are in Bennett et al. 2026.  

\section{Discussion and Conclusions}
\label{sec-discussion}
 
  Keck follow up observations of OGLE-2012-BLG-0563 yields a host mass of $M_{\rm host} = 0.81\pm 0.04\msun$, a K dwarf star, orbited by a 
Jupiter mass planet,
$M_{\rm Planet} = 1.12\pm 0.09 M_{\rm Jup}$.  This analysis of OGLE-2012-BLG-0563 shows the measured separation between lens and source is smaller than the predicted one from discovery paper \citet{fukui15}.  Now this means several options,  A) there could be a nearby unrelated star or a companion to either source or lens present in the image.  B) the model in discovery paper could be incorrect.  We found that the predictions and the light curve model in discovery paper was incorrect since it included MOA B $\&$ C data and FTS data which did not fit the model consistently in a row at about 20 data points in the same side of the model.  This indicated an obvious systematics in these two data sets.  An independent light curve modeling without these two datasets and rereduced MOA dataset yielded a different best fit light curve model.  However,  the unconstrained model with FTS and MOA B$\&$C removed still pointed towards the incorrect identification of the lens.  It was only by implementing the image constrained modeling with the lens-source relative proper motion vector that the correct lens star was detected.  This clearly demonstrates that identifying photometric systematics and implementing relative proper motion constraints are vital for accurate lens determination—a methodology that will become a powerful diagnostic tool in the upcoming Roman Space Telescope era. 
  
   The mass of the host star from current analysis is  $0.81 \msun$ and the host mass in previous \citet{fukui15} analysis was $0.34 \msun$.  The planet-host mass ratio is only slightly different, as shown in Table 5.  But this change in the host mass has resulted in a big change in planet mass.  The planet mass from the current analysis is a Jupiter mass analog planet instead of a Saturn mass planet reported in \citet{fukui15}. This change in host mass is due to the fact that the measured heliocentric relative proper $\mu_{\rm rel, H}$ from Keck analysis favors a light curve model with a smaller $t_E$ value than \citet{fukui15}.  The  $\mu_{\rm rel, H}$  provided the magnitude and direction of microlensing parallax which constrained the $\theta_E$ and the mass of the lens using Equation 5.   This also resulted in a big change in lens distance: from 1.3 kpc to 5.47 kpc. Because the primary scientific goal of the Roman Galactic Bulge Time-Domain Survey (GBTDS) is to complete the exoplanet census, variations in the calculated planetary mass and system distance significantly impact the resulting statistics. For example,  if we are studying the distribution of exoplanets across the Galactic Center, this specific planetary system's data point would shift from 1.3 kpc to 5.47 kpc.  Furthermore, if we are analyzing the mass function, systems like this will significantly alter the host mass function. This highlights the critical importance of thoroughly checking photometric systematics, as these biases can distort the derived mass and distance of a system, ultimately skewing our understanding of exoplanet demographics. 
    
  In this paper we measure a lens-source separation of $23.28 \pm 0.48$ mas in the Keck narrow images. This is the smallest separation measured in all the high angular resolution Keck observations.  Earlier to this analysis, the smallest lens-source separation measured was $\approx 29 $ mas for the event MOA-2007-BLG-192 \citep{terry2024}.  The smallest lens-source separation measured using HST data is $\approx 12 $mas for the event MOA-2010-BLG-310 from \citet{moa310}.  Being able to measure smaller separations than previous studies is very important in the context of  developing the mass measurement method for Roman.  Roman Galactic Exoplanet Survey will detect 1400+  exoplanets \citep{penny19} and observe them simultaneously with high angular resolution. The mission will observe 6 epochs of data spread across 5 years.  With an average lens host star in the galactic disc and the source in galactic bulge, the lens-source relative proper motion is about 5 mas/yr. Hence the lens-source separation expected from the Roman mission is about 25 mas for a lot of planetary events. Our analysis of OGLE-2012-BLG-563 demonstrates that a lens-source separation similar to what is expected from Roman can be detected clearly even with $\sim$20 images.  Being able to detect this small separation in Infrared passband where the PSF is fatter than the visible passbands is of high importance to Roman.  With thousands of images in F146 and a few hundreds of them in two other passbands (one red and one blue), Roman would definitely be able to detect separations like this during its prime mission.  Note that,  OGLE-2012-BLG-0563 is not in the EUCLID Galactic Bulge Survey field or Roman GBTDS field. However, it is part of Roman Galactic Plane Survey (GPS) field.  Events like OGLE-2012-BLG-0563 would verify the pipeline for Roman lens flux analysis and its limitations.  In the future,  several planetary events from the Roman GBTDS with very small separations might be followed up with Keck LGS AO for mass measurements since Keck NIRC2 narrow camera has a 10 mas pixel scale compared to Roman's 110 mas pixel.  Our paper developed the method that can be used to verify small-separation Roman microlensing planetary events with the Keck Follow-up images.
  
   Our analysis shows the importance of multi-passband image analysis and obtaining independent consistent results from Keck analysis.  \citet{moa310} has argued that not all excess flux on top of the source is necessarily lens.  For OGLE-2012-BLG-0563, the lens star is detected at consistent separation in three different passbands which confirms that this star is indeed the lens.  The lens is much fainter in HST $V$ band and brightest in $K$ band,  which made $K$ band the obvious choice for the first analysis.  After we analyzed Keck $K$ band,  we analyzed the HST $I$ band images.  The separation constraints from Keck $K$ band analysis and HST $I$ band analysis were used to constrain the lens location in HST $V$ band. Since the measured separation in high resolution images is much smaller than the predicted separation from the original discovery paper \citet{fukui15},  it was initially difficult to say if it was a nearby unrelated star or a companion to a lens or source star.  However, consistent separation measurements from three passbands and the sepation being consistent with the updated best fit light curve model shows that the identified star is indeed the lens.  The mass measurements from HST passbands yield $M_{\rm host} = 0.801 \pm 0.033 \msun$ and $M_{\rm planet}= 1.116 \pm 0.087 M _{\rm Jup}$, respectively, located at a distance of $D_{L}=5.46 \pm 0.56$ kpc.  These measurements are consistent with our measurements with Keck only image analysis,  $M_{host} = 0.81^{+0.03}_{-0.04} \msun$, $m_{p} = 1.12^{+0.08}_{-0.08} M_{\rm Jup}$ and $D_L = 5.49^{+0.58}_{-0.55}$ kpc.  For such small lens-source separations ($23.28 \pm 0.48$ mas), especially the ones not matching the original discovery paper,  multiple independent observations and measurements authenticate the result with high certainty and advance the development of mass measurement method. 
   
 The event OGLE-2012-BLG-0563 is part of S16 sample \citep{suzuki16}, hence it is one of the 29 planets in the mass ratio distribution.  As noted, in Table \ref{tab-lc}, the mass ratio of this event has increased., but it is still in the same order of magnitude.  Hence it does not change the mass ratio distribution in \citet{suzuki16} analysis.  However, the mass of the host star changes from $0.34 \msun$ in \citet{fukui15} analysis to $0.81 \msun$ in our current analysis (see Table \ref{tab-params-histogram}).  This significant increase in mass would affect the host mass function of different exoplanets, even though the mass ratio function stays intact.  The increase in host mass would influence the big fundamental questions like, which stars host which kind of planets? Hence, it is very important to check the photometric systematics and measure the correct mass.
 
 This event is also part of \citet{penny18} sample of 31 events.  \citet{penny18} argues that several events had lens located in the galactic disc, which signifies that the galactic bulge is likely to be devoid of planets. However, this is unexpected and there is likely to be some inaccuracy in the lens distance estimates.  As seen in this paper, event OGLE-2012-BLG-0563 indeed had a wrong lens distance and the correct lens distance would be $5.48^{+ 0.58}_{-0.55}$ kpc which is much closer to the bulge.  Another event from this sample, MOA-2007-BLG-192  was also estimated to be at a lens distance of $0.70^{+ 0.21}_{-0.12}$ kpc. However, a recent reanalysis by \citet{terry2024} shows that the lens distance of this event has increased to $2.16 \pm 0.30$kpc.  High angular resolution follow up analysis of these events have definitely shown an increment in the lens distance indicating that galactic bulge is not devoid of planets. 

 This work made use of data from the Astro Data Lab at
NSF’s OIR Lab, which is operated by the Association
of Universities for Research in Astronomy (AURA), Inc.
under a cooperative agreement with the National Science Foundation. We also acknowledge the help of Dr. Peter Stetson on providing us with a feedback on our analysis of Keck data. The Keck Telescope observations and analysis
was supported by a NASA Keck PI Data Award 80NSSC18K0793. Data presented 
herein were obtained at the W. M. Keck Observatory from telescope time allocated to the National Aeronautics 
and Space Administration through the agency's scientific partnership with the California Institute of Technology 
and the University of California. The Observatory was made possible by the generous financial 
support of the W. M. Keck Foundation.
DPB, AB, NK, and SKT  were also supported by NASA through grant NASA-80NSSC18K0274 and 
by NASA award number 80GSFC17M0002.
This work was supported by the University of Tasmania through the UTAS Foundation and the endowed
Warren Chair in Astronomy and the ANR COLD-
WORLDS (ANR-18-CE31-0002). This research was
also supported in part by the Australian Government
through the Australian Research Council Discovery Pro-
gram (project number 200101909) grant awarded to
Cole and Beaulieu. Work by NK is supported by JSPS KAKENHI Grant Number
JP18J00897. J.R.L. acknowledges support by the National Science Foundation under Grant No. 1909641 and the Heising-Simons Foundation under Grant No. 2022-3542. AF's work was partly supported by JSPS KAKENHI Grant Number JP17H02871. 


\end{document}